# Multiwavelength Raman spectroscopy analysis of a large sampling of disordered carbons extracted from the Tore Supra tokamak

Cédric Pardanaud[*], Céline Martin, Pascale Roubin

Aix-Marseille Univ, CNRS, PIIM UMR7345, 13397 Marseille, France

**Abstract**—Disordered carbon often exhibit a complex Raman spectrum, with four to six components. Here, a large variety of disordered carbons, forming a collection of samples with a great variety of structures, are analysed using multi-wavelength Raman microscopy (325.0, 514.5, 785.0 nm). They allow us to extend Raman behaviour known for nano-crystalline graphite to amorphous carbons, (dependence with the excitation wavelength) and other known for amorphous carbons to nano-crystalline graphite, (differentiation of the smallest cluster size probed using different excitation wavelengths). Experimental spectra were compared to simulated spectra, built using known laws, to evidence a new source of broadening.

# 1. Introduction

Raman microscopy is routinely used together with other techniques to characterize C-based materials, from nano-crystalline graphite (nc-G) to amorphous carbons, tetragonal or not (ta-C or a-C), hydrogenated or not (a-C:H or a-C). It probes the structure [1] and is highly sensitive to bonding properties by interpreting the 1000 - 1800 $cm^{-1}$ region, dominated by the G and D bands due to $sp^2$ hybridization of carbon atoms [2]. The G band is assigned to the bond stretching of both aromatic and aliphatic C-C pairs, whereas the D band (associated with a D' band) is assigned to the breathing of aromatic rings. The Raman analysis of graphite and nc-G clearly shows that this D band exists only when there is disorder [1], nc-G being most often represented as aromatic domain of limited size (typical size named $L_a$), referred to as clusters in the following. The D band has been successfully explained by a double resonance mechanism [3, 4] and, due to electronic resonance, its frequency and intensity depend on the excitation wavelength [5]. Even if the D' band's frequency do not vary drastically with the excitation wavelength, its origin was found similar to that of the D band [6]. A dependence of the D band Raman spectroscopic parameters with the excitation wavelength was also reported in detail for different types of a-C, a-C:H, ta-C, ta-C:H [7]. In these cases, the Raman effect is resonant and different cluster sizes are probed according to the excitation wavelength [8, 9], the lower wavelength, the lower cluster size. For a-C and a-C:H, the G band width is related to disorder (cluster size, cluster size distribution, chemical bonding) or stress [10]. For disordered multilayer graphene the G band width evolution is also related to disorder [11].

Spectra of a-C and nc-G are clearly distinct [7]: for example, G and D bands are much broader for a-C (width ~ 80 - 200 $cm^{-1}$) than for nc-G (width ~ 15 - 40 $cm^{-1}$). In addition their relative intensities (D band intensity over the G band intensity) strongly depend on disorder, increasing or decreasing when disorder increases, for nc-G or a-C, respectively. Note that to fit nc-G data, additional bands at ~1500 cm$^{-1}$ and at ~1200 cm$^{-1}$ are very often needed, interpreted as sp$^3$ or out-of-plane defects, or as an additional amorphous contribution [12-16], and called respectively D3 and D4 bands. Raman spectra thus contain information on disorder [10] such as the size clusters, $L_a$, [1, 8], the sp$^2$/sp$^3$ ratio, the hydrogen content [17-19], etc. Correlation with other techniques is thus needed to get quantitative information, still remaining in most cases approximate [20].

Based on a multiwavelength analysis, this paper revisits the Raman spectroscopy of nc-G and a-C, using a large variety of disordered samples which exhibit very different Raman spectra. Samples are carbon deposits extracted from the fusion device Tore Supra: they originate from different places in the machine, with different deposition and surface temperature conditions, resulting in a wide range of growth processes and provide an extensive continuous sampling over nc-G to a-C. In this paper our aim is to highlight unexpected similar Raman behaviours between both types of carbon. Samples, data acquisition and treatment are presented in section 2. Data analysis is presented in section 3. A simple model and a simulation are proposed to interpret Raman parameters in section 4 and results are discussed in section 5.

## 2. Samples and methods

2.1 Samples

We investigated here carbon materials coming from the plasma facing components of the Tore Supra tokamak, both the virgin material (C/C composite, referred to as C/C) and carbon deposits (referred to as TS) created by the interaction of the plasma ($D^+$ ions) with the plasma facing components [21, 22]. The C/C composite used in Tore Supra is a graphitic material provided by the Snecma Propulsion Solid Company, composed of carbon fibres embedded in a carbon pyrolytic matrix, whose properties are more detailed in [23]. The deposits were collected on different components inside the tokamak and were studied in detail previously, mainly by Raman microscopy and electron microscopy (nc-G [24], a-C [25, 26]). They form a wide range of carbon materials from nc-G to a-C, all described in more details in [24-26]. We used them here to get an extensive sampling over this series of materials to investigate a wide range of Raman spectra.

We also compared our Raman spectra to the Raman spectra of different types of plasma-deposited amorphous carbon layers as reference layers: ta-C:H [7], ta-C [27], a-C:H [26, 28] and ultra nano-crystalline diamond films (referred to as a-C/nc-d) which contain a $sp^2$ amorphous phase [29]. ta-C:H, ta-C and a-C:H layers were heat treated at various temperatures after deposition. The structural changes of ta-C:H occurs at ~ 450 °C: below this temperature this layer contains ~ 67 % of $sp^3$ carbon, falling at 20 % at 800 °C [30]. Conversely the ta-C, even annealed at T = 1000 °C, contains 85 % of $sp^3$ carbon, leading to a density of about 3.1 g.cm$^{-3}$. The a-C:H layer is a typical hard hydrocarbon film with a hydrogen content of 33% and a density of about 1.7 g cm$^{-3}$. The different a-C/nc-d data correspond to different synthesis conditions (for details, see [29]).

2.2 Raman microscopy

Raman spectra were recorded using a Horiba-Jobin-Yvon HR LabRAM apparatus (green excitation wavelength, $\lambda_L$= 514.5 nm, and UV excitation wavelength, $\lambda_L$=325.0 nm, with a 100X and a 40X objective, respectively) and using a Renishaw inVia apparatus (infrared excitation wavelength, $\lambda_L$=785.0 nm, with a 100X objective). The laser power was kept at less than ~ 1mW $\mu m^{-2}$ to prevent damages and spectra were recorded with various exposure times to check that samples do not evolve under irradiation. About 300 samples extracted from Tore Supra, mainly in the deposition zone (see [25, 26]) were analysed by Raman spectroscopy. Graphitic domains are small (few tens of nm, much smaller than the laser spot) and randomly oriented, and therefore orientation effect is not expected to be measured [24].

We analysed here the following Raman parameters: the G band wavenumber, $\nu_G$, the full-width at half-maximum of the G and D bands, $\Gamma_G$ and $\Gamma_D$, respectively, and the relative heights of the D and G bands ($R = H_D/H_G$). We introduced a scaled parameter by reference to the green excitation wavelength, $R_\lambda$: $R_\lambda = R \times (514.5/\lambda_L)^4$. In the case of nc-G, G and D bands are always well separated while in the case of a-C or a-C like spectra recorded using green and infrared excitations, G and D bands are broad and most of the time overlap. They can often be correctly fitted by two Gaussian bands but adding smaller components is most of the time needed. An alternative way to fit data [8] is to use an asymmetric profile for the G band, called the Breit Wigner Fano profile, and a Lorentzian profile for the D band. It is known, but frequently ignored that, depending on spectral decomposition used, the number of bands involved in the decomposition, their profile, the overlapping of the different bands, or the operator, the experimental spectral parameter values deduced can be very different: in most of the cases there is not a unique way to fit the data. To get a straightforward picture of Raman spectra, we chose next to analyse the following Raman data with the raw Raman parameters, i.e. without spectrum decomposition, measuring in case of overlap only the high-frequency and the low-frequency half-width for the G and the D bands, respectively. Figure a in supplementary information resumes how the raw spectral parameters were obtained for overlapped and not overlapped spectra (which represents two kinds of typical Raman spectra recorded with $\lambda_L$=514.5 nm). Note that this figure also gives a visual proof that Raman parameters obtained by fitting depends on the spectral decomposition chosen. As we wanted to avoid this problem, we preferred working with apparent parameters, i.e. without spectral decomposition.

Note that in a second step, presented in section 4, we have built simulated spectra, the corresponding Raman parameters of each band being chosen realistically and/or using existing relations between each other found in the literature. We have then analyzed these simulated spectra by measuring their raw Raman parameters exactly in the same way as the experimental spectra, for comparison. This method gives additional constraints compared to a classical fit which is not able to distinguish between several mathematical solutions depending on different origins listed above. Figure b of the supplementary material gives four examples of such simulated spectra.

## 3. Results

Fig. 1 displays typical Raman spectra obtained for C/C and TS samples with $\lambda_L$=325.0, 514.5 and 785.0 nm. C/C samples are graphitic carbons and their D and D' contributions are significant only for $\lambda_L$ = 785.0 nm, due to resonance effects [31]. TS spectra are more heterogeneous: some are similar to a-C spectra with broad G and D bands while some either are similar to nc-G spectra with well separated G and D bands or look intermediate, between nc-G or a-C spectra. For all samples, G and D bands both broaden and $R = H_D/H_G$ increases when excitation wavelength increases.

**Figure 1**

In Fig. 2 $\nu_G$ is plotted against $\Gamma_G$ for C/C and TS samples and for reference samples ($\lambda_L$=514.5 nm). For each type of samples, data points are remarkably aligned. C/C data points are all situated at $\Gamma_G \sim 25$ cm$^{-1}$ and $\nu_G \sim 1580$ cm$^{-1}$. TS data points continuously spread either along the positive slope straight line ($\Gamma_G$ in the range 20-80 cm$^{-1}$, $\nu_G$ in the range 1580-1600 cm$^{-1}$) or along the negative slope straight line ($\Gamma_G$ in the range 80-180 cm$^{-1}$, $\nu_G$ in the range 1600-1520 cm$^{-1}$). Part of TS data points are very close to a-C:H data points, at slightly lower frequency. This shift can be attributed to an isotopic effect as TS samples are deuterated samples (Note that the stretching mode frequency of benzene, which varies linearly with the number of bonded hydrogen and deuterium atoms, is 1599 cm$^{-1}$ for $C_6H_6$ molecules whereas it is 1557 cm$^{-1}$ for $C_6D_6$ molecules [32].). This plot shows that $\Gamma_G$ and $\nu_G$ are correlated and, as $\Gamma_G$ is usually used to estimate disorder [10], we will use only the $\Gamma_G$ parameter in what follows.

**Figure 2**

Fig. 3 displays the G band width $\Gamma_G$ as a function of the scaled intensity ratio $R_L = R \times (514.5/\lambda_L)^4$ for the three excitation wavelengths. In the case of the green excitation, C/C data points line up in the window $R = 0.1 - 0.7$ and $\Gamma_G = 17 - 30$ cm$^{-1}$: R increases when $\Gamma_G$ increases, in agreement with previous results [33]. The linear fit of these data intercepts the vertical axis at $\sim 18$ cm$^{-1}$. This latter value includes the intrinsic width due to electron phonon coupling occurring for infinite graphene planes ($\sim 11$ cm$^{-1}$ [34]) and an additional broadening of $\sim 7$ cm$^{-1}$. A few TS data points follow the linear relation found for C/C data and most of them deviate from it: R increases when $\Gamma_G$ increases in the window R = 0.7 - 1.5 and $\Gamma_G$ = 30 - 80 cm$^{-1}$. On the contrary, for more disordered carbons, R decreases when $\Gamma_G$ increases in the window R = 0.3 - 0.9 and $\Gamma_G$ = 80 - 160 cm$^{-1}$. All the green data points, from C/C to TS, i.e. from nc-G to a-C, are spread within a continuous "banana-like" cloud of points. UV and near infrared data points are also spread in similar continuous cloud, but shifted compared to the green data points. However, common trends between the three wavelengths are evidenced: this figure clearly extends the validity of the linear relation observed for green data and significantly reduces data spreading. A similar linear $\lambda_L$ dependence has been already reported in the case of nc-G for $\Gamma_G$ in the range 20 - 40 cm$^{-1}$ [31] and also in [35] and has been related to the general $\lambda_L^4$ dependence of the intensity of the G band. The plot of Fig. 3 indicates that this dependence is valid also for high $R_\lambda$ and probably also for a-C. Note that if $\Gamma_G$ had been plotted against R and not $R_L$, UV data points could appear significantly narrower than green data points, while conversely, near infrared data points could appear much more spread than green data points.

We show in what follows that a simple model relating the width and the relative intensity with the aromatic domain size $L_a$ can provide a consistent picture of the plot of Fig.3 for both nc-G, a-C and in carbons with spectra found in between nc-G and a-C.

**Figure 3**

**4. Modelling and simulation**

The Tuinstra relation [1], valid for nc-G and green excitation, is commonly used to estimate $L_a$ using:

$$R = C \, L_a^{-1} \qquad (1a)$$

with C = 4.4 nm. The authors of [31] have taken into account the $\lambda_L$ dependence which leads now to $R = C\, \varepsilon_\lambda\, L_a^{-1}$, with $\varepsilon_\lambda = (\lambda_L / 514.5)^4$. They have also shown that the G band width, $\Gamma_G$, could be approximated for such materials by a linear law:

$$\Gamma_G = B_1 + A_1\, L_a^{-1}. \qquad (2a)$$

We thus obtain a linear relation: $\qquad \Gamma_G = B_1 + A_1\, C^{-1}\, R_L, \qquad (3a)$

in agreement with data plotted in Fig. 2 for nc-G samples. A linear fit of these data leads here to $A_1 \sim 89$ nm cm$^{-1}$ and $B_1 \sim 18$ cm$^{-1}$, whereas for perfect graphite ($L_a \to \infty$) $B_1$ is determined at $\sim 11$ cm$^{-1}$ [36] which indicates the existence of an additional broadening compared to perfect graphite.

To fit a-C data, we use the relation proposed by Ferrari and Robertson [8] for R and for the green excitation wavelength:

$$R = C'\, L_a^2 \qquad (1b)$$

and we model the G band width by:

$$\Gamma_G = B_2 + A_2\, L_a^{-n}. \qquad (2b)$$

Fitting the data found in [37], which has the advantage to take into account a large variety of samples, leads to $n \sim 0.58$. Then a similar dependence with $\lambda_L$ for a-C than for nc-G has been taken into account, which looks like the one found by [35]. We also used $R_L = C'\, L_a^2$, and second, continuity between nc-G and a-C at $L_a = L_{a0}$. This finally leads to a non linear dependence:

$$\Gamma_G = B_2 + A_2\, C'^{n/2}\, R_L^{-n/2} \qquad (3b)$$

with $C' = C\, L_{a0}^{-3}$ and $A_2 = A_1\, L_{a0}^{n-1} + (B_2 - B_1)\, L_{a0}^n$.

With these simple assumptions, we succeed in fitting a-C data by only adjusting $B_2$ and $L_{a0}$ (Fig. 3), the intersection of the two fits at $R_\lambda \sim 3.6$ corresponding to $L_{a0} \sim 1.2$ nm (for a value of C=4.4 nm). Note this value of 1.2 nm has to be taken with care as it is related to the chosen value of C. Values of 2 to 3 nm were found for other samples [8, 11]. Then our $L_{a0}$ value may not be the same in other samples. The good fit obtained indicates that the $\lambda_L$ dependence of R is shared by both a-C and nc-G.

We thus divide TS samples in three groups: pure nc-G, pure a-C, and intermediate samples, these latter samples being those not correctly described by eq. 3a or eq. 3b. We propose to interpret their spectra as the sum of two sets of bands: one nc-G set composed of three bands, G, D and D', and one a-C set composed of two bands, G and D. This choice is supported by previous studies showing the need of introducing two additional broad bands (the D3 and D4 bands, see the introduction) to fit experimental data. Our method consists in first, building simulated spectra by adding these two sets of bands using realistic parameters ($\nu_G^s$, $\nu_D^s$, $\nu_{D'}^s$, $\Gamma_G^s$, $\Gamma_D^s$, $R_L^s$, ...the s-exponent standing for "simulated"), and second, analyzing these simulated spectra with their raw Raman parameters exactly in the same way as the experimental spectra.

The band shape was chosen Lorentzian for the nc-G set and Gaussian for the a-C set. Positions $\nu_G^s$, $\nu_D^s$, $\nu_{D'}^s$ were fixed to standard values [31] reported in table I, the dispersion of the nc-G D band being taken into account. The effect of a red shift of the D' band from the usual position at 1620 cm$^{-1}$ was tested and a 5 cm$^{-1}$ shift was found to have no significant effect. The scaled intensity ratio between the G and D bands, $R_L^s$, was chosen in the range of interest, i.e. from 0 to 3 for the nc-G bands and fixed at 0.6 for the a-C bands. The intensity ratio between the D' and D bands, $H_{D'}^s/H_D^s$, was fixed at 0.2. Width $\Gamma_G^s$ was then deduced from eq. 3a and 3b whereas $\Gamma_D^s/\Gamma_G^s$ was fixed at 2 for the two sets of bands (a value of 1 was also tested with no significant influence on apparent $\Gamma_G$ and $R_\lambda$ parameters) and $\Gamma_{D'}^s$ arbitrarily chosen equal to $\Gamma_D^s$. This latter hypothesis was tested to have no significant consequence in the simulation. Changes in the simulated spectra were thus investigated only by varying $R_L^s$ and the relative weight of the two sets of bands, $\eta^s$, defined here as the ratio between the G band maximum of the a-C component and that of the nc-G component. An additional broadening of the nc-G bands, equal for the G and the D bands, $\Gamma_{add}^s$, was introduced.

**Figure 4**

Fig. 4 displays experimental (points) and simulated (lines) Raman parameters in the ($R_L$, $\Gamma_G$) plot in the case of the green excitation. The curve at low $\Gamma_G$ ($\Gamma_G < 80$ cm$^{-1}$), simulation S1, is obtained by only taking into account the nc-G set of bands ($\eta^s = 0$) and by varying $R_L^s$ from 0 to 1.75. For $\Gamma_G < 30$ cm$^{-1}$ this curve corresponds to the straight line given by eq. 3a. For larger $\Gamma_G$, a deviation appears, which qualitatively reproduces the experimental trend in the window $30 < \Gamma_G < 80$ cm$^{-1}$, $0.5 < R_L < 1.4$. The origin of the deviation from a straight line is merely the apparent broadening of the G band due to the presence of the D' band. To reproduce the experimental trends at larger $\Gamma_G$, introducing the a-C set of bands is needed, as shown with simulation S2 ($\eta^s$ varying from 0 to 1 and $R_L^s$ being fixed at 1.5). Note that for $\eta^s$ values larger than 1, simulated parameters became inconsistent with experimental parameters. Finally, to correctly simulate spectra with $\Gamma_G > 80$ cm$^{-1}$ and $R_L < 1.0$, an additional broadening $\Gamma_{add}^s$ is introduced for the G and D bands of nc-G, (simulation S5: $\eta^s$ varying from 0 to 1, $R_L^s$ fixed at 1.15 and $\Gamma_{add}^s$ equal to 30 cm$^{-1}$). Fig. 4b is a zoom of Fig. 4a where the results of simulations with various $R_L^s$ and $\Gamma_{add}^s$ reproducing experimental data are reported. Remarkably, there is a linear relation between $R_L^s$ and $\Gamma_{add}^s$, $\Gamma_{add}^s$ decreases when $R_L^s$ increases (Fig 4c).

These simulations thus show that the main trends of intermediate sample data can be reproduced (i) by a single set of nc-G bands, or (ii) by adding an a-C component to this set, or (iii) by adding a band broadening to the bandwidth defined by eq.3a and 3b, or (iv) by combining (ii) and (iii). To summarize, data points following the straight line described in Fig. 3 (lower line) are from pure nc-G samples and, due to the broadening induced by the D' band, data points deviating from this line can also correspond to pure nc-G samples. Data points following the upper continuous line described in Fig. 3 correspond to pure a-C samples (not simulated here) while intermediate data points can be simulated as a combination of nc-G and a-C samples, either with or without an additional band broadening.

## 5. Discussion

We first focus on data correctly fitted by the two laws described by eq. 3.a and 3.b, i.e. pure nc-G or "pure" a-C samples. We show here that the $\lambda_L^4$ dependence of the R ratio which has been reported and explained for disordered graphene [35] and nc-G is probably also valid for a-C. Raman properties of a-C are expected to be those of the $sp^2$ / aromatic part of the material and thus it could be not surprising that a-C, as long as aromatic clusters are large enough for solid-state and quasi-periodic properties to exist, share some Raman signature with nc-G. In the case of the nc-G data correctly fitted with eq. 3a (straight line of Fig. 3), the maximum value of $R_L$ increases when the excitation wavelength decreases ($R_L$ = 0.2, 0.6, 3, for $\lambda$ = 785.0, 514.5, 325.0 nm, Fig. 3), $R_L$ increasing when $L_a$ decreases (eq. 1a): this is consistent with what is generally expected for a-C materials [38] for which the smaller laser wavelength, the smaller aromatic clusters probed, contrary to what is expected for perfect graphite, for which electronic bands cross linearly and resonance always exists [3]. To conclude this part, we have obtained strong evidences that the frontier between nc-G and a-C is less clear than expected, by revealing first, the $\lambda_L^4$ dependence of the R parameter of a-C samples, previously known for disordered graphene and nc-G samples, and second, the laser wavelength selectivity of Raman spectra according to the cluster size of nc-G samples, previously known only for a-C samples. Our results are then in agreement with results obtained on bombarded multilayer graphene [11].

Data not correctly fitted by eq. 3a and 3b correspond to more disordered carbons. Remarkably, the main experimental trends can be correctly simulated for the three excitation wavelengths used with a limited number of free parameters in addition of the relative intensity R of the D and G bands (related to $\Gamma_G$ by eq 3.a). The first parameter is the relative intensity ratio between the D and D' bands: deviation from the straight line describing nc-G (eq. 3a, Fig.3) is explained by an apparent excess of broadening of the G band due to the presence of this D' band. The second parameter is the ratio $\eta$ between an a-C component and the nc-G component. This amorphous component, composed of two bands at ~ 1500 cm$^{-1}$ and at ~ 1300 cm$^{-1}$, has been often needed to fit Raman spectra of disordered carbons [12], and correspond to the D3 and D4 bands of [14]. The third parameter is an additional broadening $\Gamma_{add}$ for the nc-G component. To better fit all the data, which is beyond the scope of this paper, adding free parameters such as the a-C component parameters (position, width, and relative intensity of the two bands), or such as an independent variation of the G and D band widths would certainly be needed. In the case of green excitation, we have found a remarkable correlation between $R_L$ and the additional broadening of the nc-G component, $\Gamma_{add}$ (Fig. 4c): $\Gamma_{add}$ increases as $R_L$ decreases, i.e. as the aromatic cluster size increases. $\Gamma_{add}$ is likely related to disorder, and consistently, disorder may be larger for large cluster than for small cluster, due to the difficulty to accommodate large aromatic areas in a disordered material. This can be related to another property, that no contribution due to disorder such as $\Gamma_{add}$ or a-C component is needed to add to reproduce UV spectra of nc-G (UV preferentially exciting small clusters). Raman spectra reveal disorder through two types of signature, (i) the D band which is associated to the famous relation between $L_a$ and the relative intensity of the D and G band, and (ii) the a-C additional component, also known as the D3 and D4 bands [14]. These two types of signatures can be related to two types of disorder. The former is attributed to the limited size of clusters, or more generally, the limited in-plane coherence of the aromatic domains [1, 33, 39], whereas the latter is often attributed to out-of-plane defects [12, 40] such as $sp^3$ hybridization influence, local curvature, or any source of off-planar defects. We have shown here that the additional broadening $\Gamma_{add}$ is most probably related to disorder similar to that evidenced by the a-C component, i.e., to out-of-plane disorder. The main difference between nc-G and a-C, beyond the mean size of aromatic clusters, can thus be represented by the way these aromatic domains are limited, i.e. with or without out-of-plane defects and roughly speaking, the quantity of in-plane disorder could thus be estimated using the R ratio whereas the quantity of the out-of-plane defects could be estimated using the $\eta$ ratio.

**Acknowledgments**

We acknowledge the Euratom-CEA association, the Fédération de Recherche FR-FCM, the EFDA European Task Force on Plasma Wall Interactions, and the French agency ANR (ANR-06-BLAN-0008 contract) for financial support. We also acknowledge W. Jacob, T. Schwarz-Selinger and C. Hopf (MPI für Plasmaphysik, Garching, Germany) for providing the a-C:H layers, B. Pégourié for providing TS samples, and A. Colombini (CICRP, Marseille, France) for assistance in experiments using the Renishaw set-up.

List of captions

**Figure 1**. Typical Raman spectra of C/C (dark) and TS (grey) samples for (a) $\lambda_L = 325.0$ nm, (b) 514.5 nm and (c) 785.0 nm. A linear background was subtracted. Spectra were normalized.

**Figure 2**. G band wavenumber, $\nu_G$, as a function of G band width, $\Gamma_G$: comparison of C/C and TS samples with reference samples (section II) ta-C:H [7], ta-C [26], a-C:H [25, 27] layers, and a-C/nc-d [28], $\lambda_L = 514.5$ nm. Straight lines are guides for the eyes.

**Figure 3**. G band width, $\Gamma_G$, as a function of the $\lambda_L$-scaled relative intensity, $R_L = R \times (514.5/\lambda_L)^4$, with $R = H_D/H_G$, for C/C and TS samples, for $\lambda_L = 325.0$, 514.5 and 785.0 nm. The inset is a zoom at low $\Gamma_G$. The grey straight line is the linear fit of nc-G data (eq. 3a) and the upper grey line is the fit of a-C data (eq. 3b).

**Figure 4**. Simulations of Raman parameters ($\lambda = 514.5$ nm). Simulation S1 is obtained with a pure nc-G component by varying $R_L^s$ from 0 to 1.75 (eq. 3a). Simulations S2 to S6 are obtained by adding to a nc-G component ($R_L^s$ = 1.50, 1.15, 1.25, 1.20, 1.05, resp.) an a-C component ($\eta^s$ = 0 to 1, squares are for $\eta^s = 0$) with the introduction of an additional broadening ($\Gamma_{add}^s$ = 0, 30, 20, 25, 40 cm$^{-1}$, resp.) (a) Same as Fig. 3 (b) zoom of Fig.4a. (c) relation between $\Gamma_{add}^s$ and $R_L^s$.

Figure 1

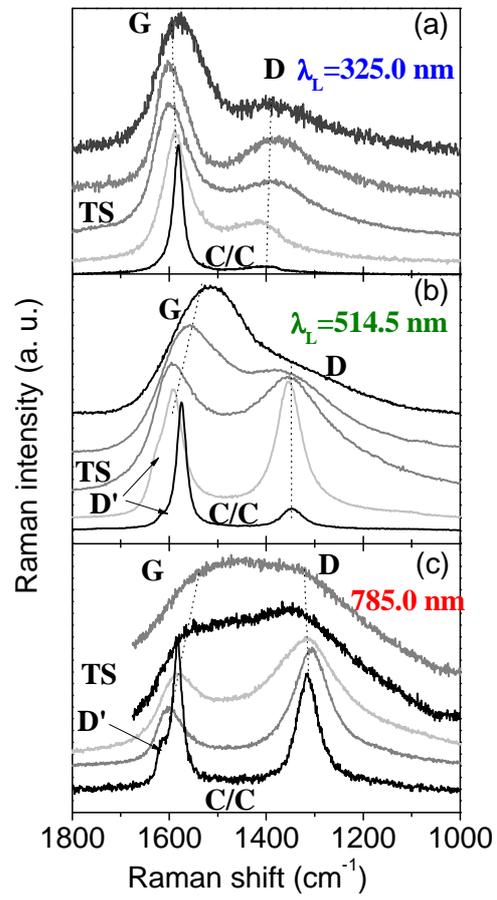

**Figure 2**

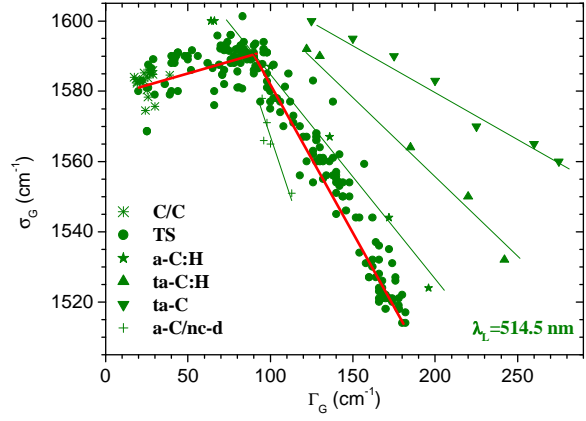

**Figure 3**

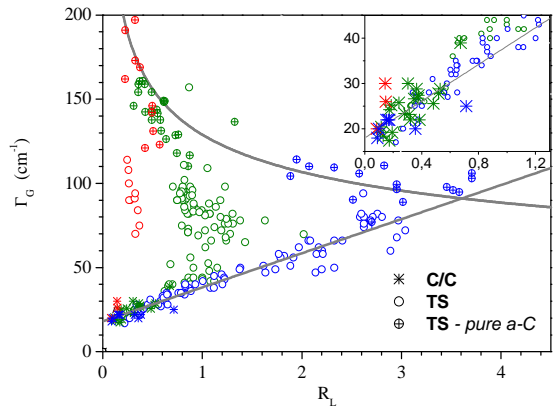

**Figure 4**

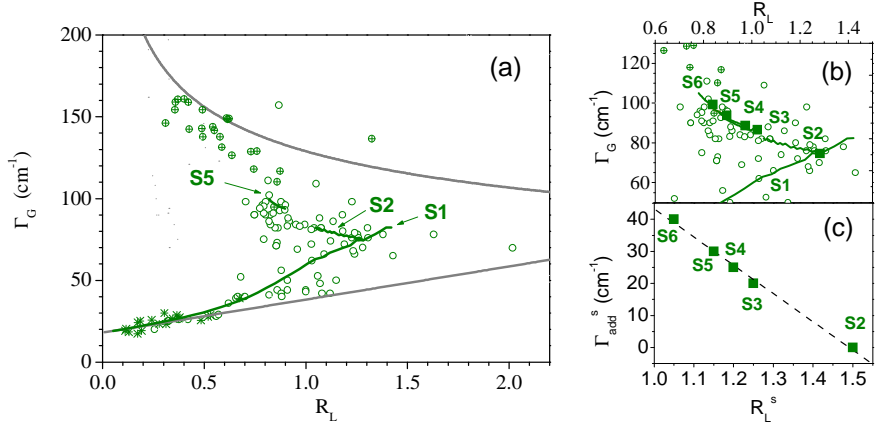

**Table 1.** Parameters of the simulated spectra for the two sets of bands: nc-G (G/D/D' bands) and a-C (G/D bands). $\lambda_L=$ [a]325.0, [b]514.5, [c]785.0 nm

|  | $\sigma_G^s$ (cm$^{-1}$) | $\sigma_D^s$ (cm$^{-1}$) | $\sigma_{D'}^s$ (cm$^{-1}$) | $H_D/H_D^s$ | $R_\lambda^s$ | $\Gamma_{D\,(D')}^s/\Gamma_G^s$ |
|---|---|---|---|---|---|---|
| nc-G | 1590 | 1420[a] | 1620 | 0.2 |  | 2 (1) |
| nc-G | 1590 | 1350[b] | 1620 | 0.2 |  | 2 (1) |
| nc-G | 1590 | 1315[c] | 1620 | 0.2 |  | 2 (1) |
| a-C | 1550 | 1300 | - | - | 0.6 | 2 (-) |

SUPPLEMENTARY INFORMATION

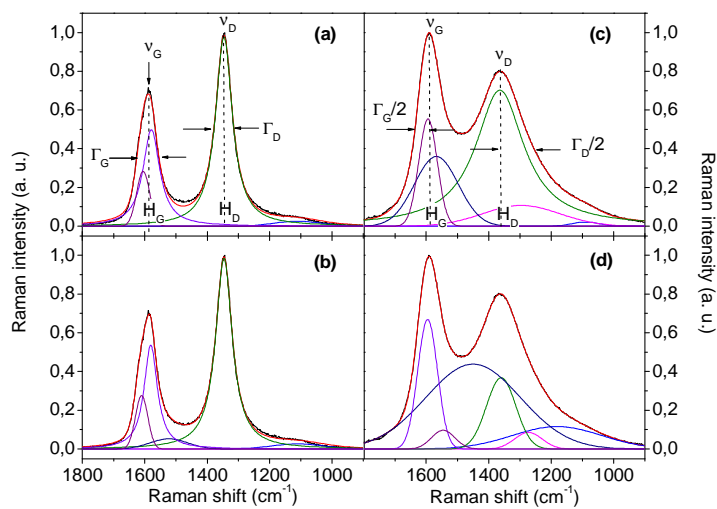

Figure a. Two kind of typical Raman spectra of Tore Supra Samples, measured using the $\lambda_L$=514.5 nm laser. (a, b) Not overlapped bands. (c, d) Overlapped bands. The raw parameters (i. e. apparent parameters) are defined in each case. Two spectral decompositions with a different number of bands have been compared in the two cases, to show that it can leads to more or less different solutions.

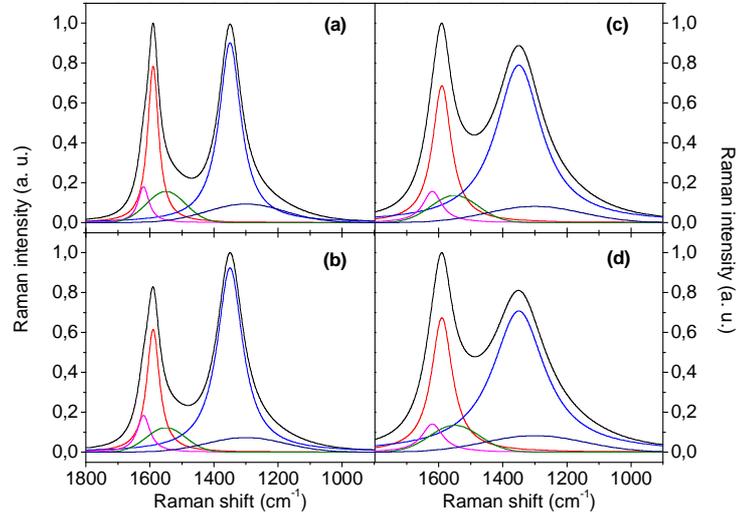

Figure b. Simulated spectra for $\lambda_L$=514.5 nm. A a-C and a nc-G component have been added. Most of the Raman spectral parameters have been fixed (see section 4 in the text for more details). The parameters that have been changed here are $R_L^s$ of the nc-G component and $\Gamma_{add}^s$. The ratio between a-C and nc-G was fixed at $\eta^s$=0.2. (a) $R_L^s = 1.15$, $\Gamma_{add}^s = 0$ cm$^{-1}$. (b) $R_L^s = 1.5$, $\Gamma_{add}^s = 0$ cm$^{-1}$. (c) $R_L^s = 1.15$, $\Gamma_{add}^s = 30$ cm$^{-1}$. (d) $R_L^s = 1.05$, $\Gamma_{add}^s = 40$ cm$^{-1}$.